\documentstyle[preprint,aps,tighten]{revtex} 
\input epsf
\newcommand{\be}{\begin{eqnarray}}
\newcommand{\ee}{\end{eqnarray}}

\def\VEV#1{\left\langle #1\right\rangle}
\def\mxth{\mathsurround=0pt }
\def\xversim#1#2{\lower2.pt\vbox{\baselineskip0pt \lineskip-.2pt
    \ialign{$\mxth#1\hfil##\hfil$\crcr#2\crcr\sim\crcr}}}

\def\bfhat#1{{\bf \hat{#1}}}
\long\def\comment#1{}

\begin{document}


\title{The Cosmic Microwave Background Bispectrum and Inflation}

\author{Limin Wang\footnote{Email:
limin@astro.columbia.edu} and Marc
Kamionkowski\footnote{Email: kamion@phys.columbia.edu}}

\address{Department of Physics, 538 West 120$^{th}$ Street,
Columbia University, New York, NY 10027}

\maketitle

\begin{abstract}
We derive an expression for the non-Gaussian 
cosmic-microwave-background (CMB) statistic $I_l^3$
defined recently by Ferreira, Magueijo, and G\'orski in terms of
the slow-roll-inflation parameters $\epsilon$ and $\eta$.  
This result shows that a
nonzero value of $I_l^3$ in COBE would rule out single-field
slow-roll inflation.  A sharp change in the slope of the inflaton potential
could increase the predicted value of $I_l^3$, but not significantly.
This further suggests that it will be difficult to account for
such a detection in multiple-field models in which density
perturbations are produced by quantum fluctuations in the scalar field
driving inflation.  An Appendix
shows how to evaluate an integral that is needed in our
calculation as well as in more general calculations of
CMB bispectra.
\end{abstract}

\section{Introduction}
Ferreira, Magueijo, and G\'orski (FMG) \cite{FMG} have recently
found evidence for a non-Gaussian distribution of
cosmic-microwave-background (CMB) temperature fluctuations in
the COBE data (as have several other groups
\cite{Pando,Novikov}).  The common lore is that this result is
inconsistent with the nearly Gaussian distribution of
temperature fluctuations expected from inflation.  Although
the result may be due to foregrounds or some curious systematic
effect \cite{Heavens,KamJaf98,BroTeg,FGM}, it is still
worthwhile to state more precisely the implications for
inflationary models if the non-Gaussianity is indeed in the CMB.
That is the purpose of this paper.\footnote{For a review of the CMB and
inflation, see \cite{annrev}.}

Inflation predicts the distribution of primordial
perturbations to be very nearly Gaussian, but self-interactions of
the inflaton field should in fact produce at least tiny
deviations from Gaussianity
\cite{SalBonBar89,Sal92,FalRanSre93,Ganetal94,Gan94}.
We derive here an analytic expression for FMG's
non-Gaussian statistic $I_l^3$  given in terms of the usual
slow-roll parameters $\epsilon$ and $\eta$ in single-field
slow-roll inflation models.  Our results verify the common expectation
that the detected value of $I_l^3$ is too large (by at
least five orders of magnitude!) to be consistent with slow-roll
inflation.  Motivated by evidence for a break in the power
spectrum of the galaxy distribution
\cite{Einasto,Peacock,Gaztanaga}, it is natural to consider
inflation models in which the slope of the inflaton potential
has a discontinuity \cite{Starobinsky,Lesg}.  We show that
the increase of the predicted non-Gaussian signal is insignificant.
We infer from this that such
a large value of $I_l^3$ should also be hard to come by in
multiple-field models in which density perturbations
are produced by quantum fluctuations in the field driving
inflation.  (Counter-examples include the models discussed in
Refs. \cite{AllGriWis87,Fanbardeen,Peebles}.)

We begin by reviewing in Section II the calculation of the
large-angle CMB power spectrum.  We then move on in Section III 
to the calculation of the CMB bispectrum and present our result
for $I_l^3$ in Section IV.  Section V considers an inflation
model with a discontinuity in the slope of the inflaton
potential.  Section VI provides a
discussion.  An Appendix provides a recursive technique for
evaluating an integral involving the product of three spherical
Bessel functions that is needed for our calculation and will be
needed for more general calculations of CMB bispectra.

\section{Two-point correlation function and power spectrum}
We start by reviewing the calculation of the angular two-point 
correlation function and power spectrum of CMB temperature fluctuations.
The photon temperature perturbation at a spacetime point
$({\bf x},\tau)$ can be Fourier expanded:
\begin{equation}  \label{define}
\frac{\Delta T}{T}({\bf x},\tau,\bfhat{n})  
 =  \int{d^3 k\, e^{i{\bf k}\cdot {\bf x}}\,  \Delta ({\bf k},\bfhat{n},\tau )},
\end{equation}
where $\bfhat{n}$ is the direction of photon momentum.
We always set the observation point to be at the origin
(${\bf x}=0$) and at the present epoch ($\tau=\tau_0$),
so we will not explicitly write these two variables in the following
derivations.
By Legendre expansion, we have
\begin{equation}  \label{Delta}
\frac{\Delta T}{T}(\bfhat{n}) =\int{d^3k\, \Delta ({\bf k},\bfhat{n})}
=\int{d^3k \sum^{\infty}_{l=0}{(-i)^l(2l+1)\psi({\bf k})\Delta_l(k)
P_l(\bfhat{k}\cdot \bfhat{n})}},
\end{equation}
where $\psi({\bf k})$ is the initial gravitational-potential perturbation
and $\Delta_l(k)$ is the photon transfer function.  We have used
the fact that the photon evolution equation is independent of 
the wavevector direction $\bfhat{k}$.  For a stationary random
process, we have
\be  \label{p2}
\langle \psi({\bf k_1}) \psi({\bf k_2}) \rangle
=P^{(2)}_{\psi}(k)\delta_D({\bf k}_1+{\bf k}_2),
\ee
where the amplitude $P^{(2)}_{\psi}(k)$ is the primordial power
spectrum.  For a scale-free primordial power spectrum, we have
\be
P^{(2)}_{\psi}(k) \propto k^{n-4},
\ee
where index $n=1$ corresponds to a flat scale-invariant
spectrum, which is close to those favored by generic
inflationary models.  The CMB temperature pattern may be written 
in a spherical-harmonic expansion with coefficients,
\be
     a_{lm} = \int \, d\bfhat{n} \, Y_{lm}(\bfhat{n}) \, \frac{\Delta
     T}{T}(\bfhat{n}).
\label{almdefn}
\ee
The angular two-point correlation function is
\be \label{twop}
\xi(\bfhat{n}_1,\bfhat{n}_2)
&\equiv& \VEV{ \frac{\Delta T}{T}(\bfhat{n}_1) \frac{\Delta T}{T}(\bfhat{n}_2) 
} =\sum_m \frac{2l+1}{4\pi}\, C_l \, P_l(\bfhat{n}_1 \cdot \bfhat{n}_2).
\ee
where
\be C_l = (4\pi)^2\int{k^2 \, dk \, P^{(2)}_{\psi}(k) \, 
|\Delta_{l}(k)|^2}
\label{eq:Cls}
\ee
is the CMB power spectrum.  We have used Eqs.~(\ref{Delta}) and
(\ref{p2}) to obtain Eq. (\ref{twop}), and the
spherical-harmonic addition theorem, orthonormality of spherical
harmonics, and 
\be \label{cl}
\VEV{a_{l_1 m_1}a_{l_2 m_2}^*}
=\delta_{l_1 l_2}\delta_{m_1 m_2} \, C_l,
\ee
to obtain Eq. (\ref{eq:Cls}).

\section{Three-point correlation function and bispectrum}
If the primordial random density fluctuation is non-Gaussian,
then the three-point correlation function is in general
non-vanishing.  We then have
\be \label{p3}
\langle \psi({\bf k_1}) \psi({\bf k_2}) \psi({\bf k_3}) \rangle
=P^{(3)}_{\psi}(k_1,k_2,k_3)\delta_D({\bf k}_1+{\bf k}_2+{\bf k}_3),
\ee
where
$P^{(3)}_{\psi}(k_1,k_2,k_3)$ 
is the spatial bispectrum of the gravitational potential.
The angular three-point correlation function for the CMB can be
written as
\be \label{threep}
\xi(\bfhat{n}_1,\bfhat{n}_2,\bfhat{n}_3) 
&\equiv& \VEV{ \frac{\Delta T}{T}(\bfhat{n}_1) 
\frac{\Delta T}{T}(\bfhat{n}_2) 
\frac{\Delta T}{T}(\bfhat{n}_3)}
\\ \nonumber
&=& \sum_{l_i,m_i} \langle a_{l_1 m_1}a_{l_2 m_2}a_{l_3 m_3} \rangle
Y_{l_1 m_1}(\bfhat{n}_1)Y_{l_2 m_2}(\bfhat{n}_2) Y_{l_3 m_3}(\bfhat{n}_3),
\ee
where
\be \label{bispectrum}
\langle a_{l_1 m_1}a_{l_2 m_2}a_{l_3 m_3} \rangle 
&=& (4\pi)^3 (-i)^{l_1+l_2+l_3}
\int{d^3k_1 \, d^3k_2 \, d^3k_3} 
\\ \nonumber & & \times Y_{l_1 m_1}^*(\bfhat{k}_1)Y_{l_2 m_2}^*(\bfhat{k}_2)
Y_{l_3 m_3}^*(\bfhat{k}_3) \delta_D({\bf k}_1+{\bf k}_2+{\bf k}_3) 
\\ \nonumber
&&
\times
P_{\psi}^{(3)}(k_1,k_2,k_3)\Delta_{l_1}(k_1)\Delta_{l_2}(k_2)\Delta_{l_3}(k_3).
\ee
This last equation is obtained by using Eqs. (\ref{almdefn}),
(\ref{Delta}), and the spherical-harmonic addition theorem.
By using
\be  \label{delta}
\delta_D ({\bf k_1}+{\bf k_2}+{\bf k_3})
=\frac{1}{(2\pi)^3} \int^{\infty}_{-\infty}
{e^{i({\bf k}_1+{\bf k}_2+{\bf k}_3)\cdot {\bf x}} d^3x},
\ee
\be \label{exp}
e^{i{\bf k}\cdot{\bf x}}=4\pi\sum_{l}i^lj_l(k x)
\sum_m Y_{lm}(\bfhat{k})Y_{lm}^*(\bfhat{x}),
\ee
and the Gaunt integral,
\be \label{threeY}
\int{d\Omega \,Y_{l_1 m_1}Y_{l_2 m_2}Y_{l_3 m_3}}
=\sqrt{\frac{(2l_1+1)(2l_2+1)(2l_3+1)}{4\pi}}
\left( \begin{array}{ccc} l_1 & l_2 & l_3 \\
		 0   & 0   & 0 \end{array} \right)
\left( \begin{array}{ccc} l_1 & l_2 & l_3 \\
		 m_1   & m_2   & m_3 \end{array} \right),
\ee
where the (...) is the Wigner 3$j$ symbol, we obtain
\be \label{b3gen2}
\langle a_{l_1 m_1}a_{l_2 m_2}a_{l_3 m_3} \rangle 
= \left( \begin{array}{ccc} l_1 & l_2 & l_3 \\
		 m_1   & m_2   & m_3 \end{array} \right)
B_{l_1 l_2 l_3},
\ee
where
\be \label{b3gen4}
B_{l_1 l_2 l_3} &=&(8\pi )^3 \sqrt{\frac{(2l_1+1)(2l_2+1)(2l_3+1)}{4\pi}}
\left( \begin{array}{ccc} l_1 & l_2 & l_3 \\
		 0   & 0   & 0 \end{array} \right)  \\ \nonumber
&& \times \int{k_1^2 \, dk_1 \, k_2^2\, dk_2\, k_3^2\, dk_3}\,
J_{l_1 l_2 l_3}(k_1,k_2,k_3)
P_{\psi}^{(3)}(k_1,k_2,k_3)\Delta_{l_1}(k_1)\Delta_{l_2}(k_2)\Delta_{l_3}(k_3)
\ee
is the CMB bispectrum.
The integral,
\be \label{3Bessel}
J_{l_1 l_2 l_3}(k_1,k_2,k_3)&=&
\int{j_{l_1}(k_1 x)\, j_{l_2}(k_2 x)\, j_{l_3}(k_3x)\, x^2\, dx}
\ee
can be calculated relatively quickly by using the recurrence relations
discussed in the Appendix.
Eq.~(\ref{b3gen4}) provides a general formalism for calculating
the bispectrum of the CMB starting from any primordial spatial
bispectrum.

FMG defined a statistic $\hat{B_l}$ which is related to the
bispectrum by
\be  \label{hatB}
\hat{B_l}&=&\frac{1}{(2l+1)^{3/2}}
\left(\begin{array}{ccc} l & l & l \\ 0 & 0& 0 \end{array} \right)^{-1}
B_{lll},
\ee
and this is especially easy to calculate from Eq.~(\ref{b3gen4}).
Because of the symmetry, we can assume $k_1<k_2<k_3$ and introduce
three parameters, $r$, $u$, and $v$, defined by, 
\be
k_1 &=& r, \\
k_2 &=& ur, \\
k_3 &=& (u+v)r,
\ee
where $u>1$ and $0<v<1$, to ensure the triangle relation between
the three $k$'s.  Finally, we get
\be  \nonumber
\hat{B_l}&=&6\frac{(8\pi)^3}{\sqrt{4\pi}}
\int_0^{\infty} r^8 \, \Delta_l(r) \, dr 
\int_1^{\infty}u^2 \, \Delta_l(ur) \, du \\ 
&&\times \int_0^1(u+v)^2 \, dv\,  J_{lll}\Bigl(r,ur,r(u+v)\Bigr) 
P_{\psi}^{(3)}\Bigl(r,ur,r(u+v)\Bigr)
\Delta_l\Bigl(r(u+v)\Bigr).
\ee
The factor 6 comes from permutation of $k_1$, $k_2$, and $k_3$.

At the large angular scales relevant for COBE, the Sachs-Wolfe
effect \cite{SW} dominates, and we simply have
$\Delta_l(k) = A_{\rm sw} j_l(k\Delta\eta)$, where
$\Delta\eta=(\eta_0-\eta_*)$ is the conformal time between now
and the surface of last scatter, and $A_{\rm sw}=1/3$ for a
critical-density model with primordial adiabatic perturbations.

\section{Single-Field Slow-roll inflation}
The formalism outlined above is general and applicable
to any form of $P_{\psi}^{(3)}$.  However, as pointed out by Luo\cite{luo},
the $P_{\psi}^{(3)}$ generated by slow-roll inflation models
\cite{FalRanSre93} has a special form that simplifies the calculation.
In fact, the power-spectrum normalized bispectrum,
\be  \label{I3def}
I_l^3\equiv \frac{\bfhat{B}_l}{(C_l)^{3/2}},
\ee 
that was extracted from the COBE-DMR data by FMG
can be expressed in a very simple form for models with the following form of
the spatial three-point correlation function:
\be
P_{\psi}^{(3)}(k_1,k_2,k_3)=f(k_1,k_2)+f(k_2,k_3)+f(k_1,k_3),
\ee
where $f(x,y)$ is an arbitrary function of $x$ and $y$. We start with
the Sachs-Wolfe formula,
\be \label{SW}
\frac{\Delta T}{T}(\bfhat{n})=A_{\rm sw}
\int{\, d^3k\,  \psi({\bf k})\, e^{i\Delta\eta {\bf k}\cdot \bfhat{n}}}.
\ee
In this case, the angular three-point correlation function becomes
\be \label{SWtp}  \nonumber
\xi(\bfhat{n}_1,\bfhat{n}_2,\bfhat{n}_3) 
&=&A_{\rm sw}^3
\int{d^3k_1\, d^3k_2\, d^3k_3\, e^{i\Delta\eta({\bf k}_1\cdot \bfhat{n}_1
+{\bf k}_2\cdot \bfhat{n}_2+{\bf k}_3\cdot \bfhat{n}_3)}} \\ 
&& \times \delta_D({\bf k}_1+{\bf k}_2+{\bf k}_3)[f(k_1,k_2)
+f(k_2,k_3)+f(k_1,k_3)].
\ee
Now, let us calculate the first term associated with $f(k_1,k_2)$.  By
integrating out $k_3$, then using spherical-harmonic
orthonormality, Eqs.~(\ref{exp}) and (\ref{threeY}), and
\be \label{twoY}
\sum_{lm}Y_{lm}(\bfhat{n}_1)Y_{lm}^*(\bfhat{n}_2)
=\delta_D(\bfhat{n}_1-\bfhat{n}_2),
\ee
we have the bispectrum from the first term of Eq.~(\ref{SWtp}),
\be \label{B3_1}    \nonumber
B_{l_1 l_2 l_3}({\rm first\, term})&=&
A_{\rm sw}^3(4\pi)^4
\sqrt{\frac{(2l_1+1)(2l_2+1)(2l_3+1)}{4\pi}}
\left( \begin{array}{ccc} l_1 & l_2 & l_3 \\
		0   & 0   & 0 \end{array} \right) \\
&&\times \int{dk_1\, dk_2\, k_1^2\, k_2^2\,  f(k_1,k_2) } 
j_{l_1}^2(k_1\Delta\eta)j_{l_2}^2(k_2\Delta\eta),
\ee
and similarly for the second and third terms.
The quantity $I_l^3$ then takes the form,
\be
I_l^3=\sqrt{4\pi}\frac{3\int{dk_1\, dk_2\, k_1^2\, k_2^2 \,f(k_1,k_2)\, } 
j_{l}^2(k_1\Delta\eta)\, j_{l}^2(k_2\Delta\eta)}
{\left[\int{dk \, k^2\, P_{\psi}^{(2)}(k)\,
j_{l}^2(k\Delta\eta)}\right]^{3/2}}.
\ee
For slow-roll-inflation models, $f(k_1,k_2)$ takes the form \cite{FalRanSre93,Ganetal94}
\be  \label{slowroll}
f(k_1,k_2)=A_{\rm infl}P_{\psi}^{(2)}(k_1)P_{\psi}^{(2)}(k_2),
\ee
where 
\be \label{Ainfl}
     A_{\rm infl} = \frac{25\, m_{\rm Pl}^2}{48\,\pi} \left [ \left(
     \frac{V'}{V} \right)^2 - \frac{4\, V''}{5\, V} \right]
\ee
is a constant depending on the height $V$ of the inflaton
potential and its first and second derivatives, $V'$ and $V''$,
respectively, with respect to the inflaton $\phi$ \cite{Ganetal94}.
Thus,
\be
I_l^3=3\sqrt{4\pi}A_{\rm infl}
\left[\int{dk\,  k^2\, P_{\psi}^{(2)}(k)\,
j_{l}^2(k\Delta\eta)}\right]^{1/2}.
\ee
For a flat scale-invariant spectrum of primordial density
perturbations, $P_{\psi}^{(2)}(k) = A_{H} k^{-3}$, where (using
our Fourier and scale-factor conventions),
\be
A_{H} = \frac{96\,V^3}{25\, (V')^2\, m_{\rm Pl}^6}.
\ee
We thus find that in terms of the usual slow-roll parameters
\cite{annrev},
\begin{equation}
     \epsilon\equiv {m_{\rm Pl}^2 \over 16 \pi} \left({ V' \over 
     V} \right)^2 , \ \ \ \ 
     \eta \equiv {m_{\rm Pl}^2 \over 8\pi} \left[{ V'' \over V}
     - {1 \over 2} \left({V' \over V} \right)^2 \right],
\label{eq:epsilon}
\end{equation}
our central result for $I_l^3$ can be expressed as
\be
\sqrt{l(l+1)}I_l^3 = {2 \over m_{\rm Pl}^2} \sqrt{ 3V
     \over \epsilon} (3\epsilon -2\eta).
\label{eq:central}
\ee

Given that $(V^{1/2}/m_{\rm Pl}^2)/\sqrt{\epsilon} \sim H^2 /
\dot \phi \sim 10^{-5}$ is the density-perturbation amplitude,
and $\epsilon,\eta \lesssim 1$ in slow-roll inflation, it is
clear that a value $\sqrt{l(l+1)} I_l^3\sim1$ is inconsistent
with single-field slow-roll inflation.

\section{Single-Field Model with a Feature}

Eqs.~(\ref{eq:epsilon}) and (\ref{eq:central}) suggest that the
bispectrum could be larger if the inflaton potential was not as
smooth as required for slow-roll inflation.  Thus, 
consider now an inflation model with a ``feature'' in
the inflaton potential, such as in those models introduced to
account for the claimed detection of a feature in the measured
mass power spectrum \cite{Einasto,Lesg,Unpub}.
More precisely, consider a single-field inflation model in which the
inflaton potential is continuous, but its scaled slope $\delta\equiv 
m_{\rm Pl}(V'/V) = 4(\pi\epsilon)^{1/2}$ is discontinuous at
$k_0$.  The derivative of the
scaled slope (with respect to the field $\phi$) is proportional
to a Dirac delta function, $\delta'= A_{\rm feat} k_0 m_{\rm Pl}^{-1}
\delta_D(k-k_0)$.  The spike occurs at the value that
$\phi$ has when the comoving wavenumber $k_0$ exits the
horizon, and $A_{\rm feat}$ is the dimensionless amplitude of
the delta function.

We must have $\delta\lesssim1$ on both sides of the delta
function, and this places a constraint to the amplitude $A_{\rm
feat}$.  The change to $\delta$ as the field passes through
the Dirac delta function is
\be
     \Delta\delta = \int\, \delta' \,d\phi = \int \frac{A_{\rm feat} 
     k_0}{m_{\rm Pl}} \delta(k-k_0) \, d\phi,
\ee
and this must be small compared with unity.
Using $d\phi=(d\phi/d\ln k)d\ln k \simeq (\dot\phi/H)(dk/k)$ and 
$\dot\phi/H \simeq (m_{\rm Pl}^2/8\pi)(V'/V)$ during slow-roll,
we get $\Delta\delta \simeq A_{\rm feat}\delta/8\pi$, and so
$A_{\rm feat} \lesssim 8 \pi$.  

To proceed more precisely, we must realize that the derivation
\cite{FalRanSre93,Ganetal94} that leads to our
Eq. (\ref{slowroll}) is valid only
if the inflaton potential varies smoothly.  By including
Dirac delta function in Eq. (32) in Ref. \cite{Ganetal94}, we arrive
at the correct expression for the spatial
gravitational-potential bispectrum for an inflaton potential
with a feature:
\be
P_{\psi}^{(3)}(k_1,k_2,k_3)=\frac{5}{6(2\pi)^4}A_{\rm feat}k_0\delta(k_1-k_0)
P_{\psi}^{(2)}(k_2)P_{\psi}^{(2)}(k_3) + \{k_1 \leftrightarrow k_2\}
+ \{k_1 \leftrightarrow k_3\}.
\ee
We can still approximate $P_{\psi}^{(2)} = A_H k^{-3}$ (since we 
are still considering only $\epsilon,\eta \ll 1$ other than the point
$k=k_0$), and then from
Eqs.~(\ref{b3gen4}) and (\ref{hatB}), we have
\be  \nonumber
\bfhat{B}_l=&& \frac{40}{3\pi^{3/2}}A_{\rm feat}A_H^2
\int{k_1^2\, dk_1 \, k_2^2 \, dk_2 \, k_3^2\, dk_3}\, J_{l\, l\,
l}(k_1,k_2,k_3)
\Delta_{l}(k_1)\Delta_{l}(k_2)\Delta_{l}(k_3)
\frac{k_0\delta (k_1-k_0)}{k_2^3 \,k_3^3}  \\
&&
+ \{k_1 \leftrightarrow k_2\}
+ \{k_1 \leftrightarrow k_3\}.
\ee
By the symmetry of $k_1$,$k_2$, and $k_3$, we obtain
\be \nonumber
I_l^3=&& \left( \frac{H}{m_{\rm Pl}\delta}\right) \frac{6}{\pi^2}A_{\rm feat}
{\tilde k}_0^3 \Delta_l({\tilde k}_0)
\left( \int_0^{\infty}{|\Delta_l({\tilde k})|^2{\tilde k}^{-1}
d{\tilde k}} \right)^{-3/2} \\
&& \times \int_0^{\infty}{\Delta_l({\tilde k}_2){\tilde k}_2^{-1} d{\tilde k}_2
\int_{|{\tilde k}_0 - {\tilde k}_2|}^{{\tilde k}_0+{\tilde k}_2}
{\Delta_l({\tilde k}_3)j_{lll}({\tilde k}_0,{\tilde k}_2,{\tilde k}_3)
{\tilde k}_3^{-1} d{\tilde k}_3}},
\ee
where ${\tilde k}_x=k_x \Delta \eta$.  Again, during slow roll,
the first term $H/(m_{\rm Pl}\delta)
\sim H^2/\dot{\phi} \sim 10^{-5}$.  Numerical integration shows that
the ${\rm rest}$ expression peaks at a value of $l$ corresponding to
the scale of $k_0$, but with an amplitude $\lesssim 0.1\,A_{\rm 
feat}$.  Therefore, even with a discontinuity in the slope of the
inflaton potential, the predicted $I_l^3$ is still several orders
of magnitude smaller than unity (the desired value to explain the
COBE anomaly found by FMG around $l=16$)
as long as the slow-roll conditions are satisfied everywhere else. 

\section{Discussion}
We gave a numerically manageable general formalism that allows us to 
calculate the bispectrum of the CMB starting from an arbitrary spatial
bispectrum.  For certain forms of the
spatial bispectrum, including those given
by single-field slow-roll inflation models, the Sachs-Wolfe
calculation results in an analytic expression for the
non-Gaussian statistic $I_l^3$; it turns out to be roughly the
density-perturbation amplitude times a linear combination of the
slow-roll parameters $\epsilon$ and $\eta$.  This result
demonstrates that FMG's non-Gaussian signal is
inconsistent with single-field slow-roll inflation models.  The
predicted value of $I_l^3$ can be increased if the slope of the
inflaton potential is discontinuous, but even this larger
non-Gaussian signal is too small to account for the detection
found in Ref. \cite{FMG}.  Although a complete calculation
would require detailed specification of a model, our results for
the discontinuous single-field model suggest that generic
multiple-field models designed to produce a break in
the slope of the mass power spectrum \cite{Starobinsky,Lesg}
will be unable to produce a non-Gaussian signal large enough to
explain the COBE result.  Counter-examples can be
found.  In particular, inflation models can be designed in
which density perturbations are produced by quantum fluctuations 
of some scalar field other than that driving inflation.  For
example, in Ref. \cite{AllGriWis87}, non-Gaussian isocurvature
perturbations to an axion density are produced by quantum
fluctuations in the axion field, but inflation is driven by some 
other scalar field not associated with Peccei-Quinn symmetry
breaking.  The mechanism of Ref. \cite{Peebles} is somewhat
similar.  In the model of Ref. \cite{Fanbardeen}, non-Gaussian
density perturbations are produced (at least in part) by the
square of a second scalar field (although details of the
inflationary dynamics are not presented).  
In conclusion, if the detection of nonzero
$I_l^3$ is ultimately attributed to the CMB, it raises a serious
problem for inflationary models where quantum fluctuations in 
the inflaton give rise to large-scale structure.

\acknowledgments
We thank R. Friedberg for suggesting the recursive technique
for evaluating integrals with oscillatory integrands.
We thank K. G\'orski, A. Kosowsky, J. Lesgourges, J. Magueijo, 
and L. Verde for useful comments and discussions.
This work was supported by a DoE Outstanding Junior Investigator
Award, DE-FG02-92ER40699, NASA grant NAG5-3091, and the Alfred
P. Sloan Foundation. 

\section*{appendix}
We need to evaluate the following integral:  
\be   \label{theintegral}
J_{l_1 l_2 l_3}(k_1,k_2,k_3)
=\int_{0}^{\infty} \, j_{l_1}(k_1 x) \, j_{l_2}(k_2 x) \,
j_{l_3}(k_3 x)\, x^2\, dx ,
\ee
where $j_l(x)$ is spherical Bessel function.  The wavenumbers
$k_1$, $k_2$, and $k_3$ satisfy the triangle relation (they
should be able to form a triangle), and $l_1+l_2+l_3$ has to be
an even number.  To evaluate the integral, we use the recursive technique
discussed in Ref. \cite{Friedberg}.  To begin, we evaluate the
integral for some special cases:
\be  \nonumber  
J_{l\, 0\, 0}
&=&\int_0^{\infty}{\frac{1}{2 i^l}\int_{-1}^{1}P_l(u) \, e^{iu
k_1 x} \, du
\frac{\sin{k_2 x}}{k_2 x}\, \frac{\sin{k_3 x}}{k_3 x} x^2 dx}\\ \nonumber
&\stackrel{l\, {\rm even}}{=}&
{\frac{1}{2i^l}\frac{1}{k_2k_3}\int_{-1}^{1}P_l(u)\, du
\int_{-\infty}^{\infty}e^{iu k_1 x}\, \sin{k_2 x}}\, \sin{k_3 x} 
\, dx\\  \nonumber
&=&\frac{-\pi}{4i^lk_2k_3}
\int_0^1 P_l(u)\, du\, [\delta(uk_1+k_2+k_3)-\delta(uk_1-k_2+k_3)- \\ \nonumber
&& \delta(uk_1+k_2-k_3)+\delta(uk_1-k_2-k_3)] \\ \nonumber
&\stackrel{\Delta\, {\rm relation}}{=}&
\frac{\pi}{4i^lk_2k_3}\int_0^1 P_l(u)du\delta(uk_1-|k_2-k_3|) \\
&=&\frac{\pi}{4i^l}\frac{1}{k_1k_2k_3}P_l\left(\frac{k_2-k_3}{k_1}\right);
 \label{l00}
\ee
\be  \nonumber  
J_{l, -1, 0}
&=&\int_0^{\infty}{\frac{1}{2 i^l}\int_{-1}^{1}P_l(u) \, e^{iu
k_1 x}\, du
\frac{\cos{k_2 x}}{k_2 x}\, \frac{\sin{k_3 x}}{k_3 x} x^2 dx}\\ \nonumber
&\stackrel{l\, {\rm odd}}{=}&
{\frac{1}{2i^l}\frac{1}{k_2k_3}\int_{-1}^{1}P_l(u)du
\int_{-\infty}^{\infty}e^{iu k_1 x}\, \cos{k_2 x}} \, \sin{k_3
x}\,  dx\\  \nonumber
&=&\frac{\pi}{4i^{l+1}k_2k_3}
\int_0^1 P_l(u)du[\delta(uk_1+k_2+k_3)+\delta(uk_1-k_2+k_3)- \\ \nonumber
&& \delta(uk_1+k_2-k_3)-\delta(uk_1-k_2-k_3)] \\ \nonumber
&\stackrel{\Delta\, {\rm relation}}{=}&
\frac{\pi}{4i^{l+1}k_2k_3}\int_0^1 P_l(u)du[\delta(uk_1-k_2+k_3) 
-\delta(uk_1+k_2-k_3)] \\
&=&\frac{\pi}{4i^{l+1}}\frac{1}{k_1k_2k_3}P_l\left(\frac{k_2-k_3}{k_1}\right);
 \label{l-10}
\ee
\be  \label{l-1-1}
J_{l,-1, -1} = 
\frac{\pi}{4i^l}\frac{1}{k_1k_2k_3}P_l\left(\frac{k_2-k_3}{k_1}\right).
\ee
To evaluate the integral for other cases, we use the recursion relation,
\be
j_{l-1}(\alpha)+j_{l+1}(\alpha)=\frac{2l+1}{\alpha}j_l(\alpha),
\ee
to obtain
\be \nonumber
D_{l_1 l_2 l_3}(k_1,k_2,k_3)
&=&\int_0^{\infty}\, j_{l_1}(k_1 x)\, j_{l_2}(k_2 x)\,
j_{l_3}(k_3 x)\, x \,dx \\ 
&=&\frac{k_1}{2l_1+1}(J_{l_1-1, l_2, l_3} + J_{l_1+1, l_2, l_3}) \\
&=&\frac{k_2}{2l_2+1}(J_{l_1, l_2-1, l_3} + J_{l_1, l_2+1, l_3}) .
\ee
Therefore, we have the recursion relation we need:
\be  \label{recur}
J_{l_1, l_2+1, l_3} = \frac{k_1}{k_2}\frac{2l_2+1}{2l_1+1}
(J_{l_1-1, l_2, l_3} + J_{l_1+1, l_2, l_3})-J_{l_1, l_2-1, l_3}.
\ee
If we begin with Eqs.~(\ref{l00}), (\ref{l-10}), and
(\ref{l-1-1}), then we can use Eq. (\ref{recur}), to
recursively evaluate the integral (\ref{theintegral}) for any
combination of $l_1$, $l_2$, and $l_3$.

\newpage

\end{document}